\documentclass[12pt]{JHEP3}

\usepackage{amsmath,cite}

\title{Several new actions of $p$-branes based on bulk scalar fields}

\author{Yi-Shi Duan\\
Institute of Theoretical Physics, Lanzhou University, Lanzhou
730000, P. R. China}

\author{Zhen-Bin Cao\\
Institute of Theoretical Physics, Lanzhou University, Lanzhou
730000, P. R. China\\
Email: \email{caozhb04@lzu.cn}}

\abstract{Generally, $p$-branes play central roles in revealing the
nonperturbative structures of the string/M-theory. In this paper,
through a topological tensor current constructed in terms of a bulk
scalar field, we first show some topological properties of
$p$-branes, and then mainly discuss the construction of several new
actions of $p$-branes. We show that the actions we construct which
can be reduced to the Nambu-Goto action or some other simple
actions, are all defined naturally in the bulk spacetime, preserving
the full spacetime Lorentz invariance and satisfying a self-dual
condition. }


\begin{document}

\section{Introduction}                                                          

A new idea that we may live on a brane, namely the so-called
`brane-world' scenario of cosmology based on string/M-theory, has
been largely discussed in the last several years. $Dp$-branes, and
for more general cases, $p$-branes, which are extended objects
embedded in a higher dimensional bulk spacetime, have been found to
play important roles in revealing the nonperturbative structures of
the modern string/M-theory \cite{Polchinski-string-book,
bbs-string-book,Polchinski-brane} and so that received much
attention. As they have also been proved to be topological defects
in gauge theory \cite{Duff95}, we can study them by using some
proper topological methods.

For $p$-branes, besides various antisymmetric tensor gauge fields
living on them, in the context of the effective $d=10$ or $11$
supergravity theory they are themselves $p$-dimensional extended
sources for a $(p+2)$-form gauge field strength in the bulk
spacetime. Meanwhile, in the string/M-theory context one could also
expect some other scalar fields, associated with many moduli fields,
as fundamental fields, which, in principle, propagate in the bulk
spacetime. For example, the creation of a brane world with a bulk
scalar field, the effective 4-dimensional Einstein equations on the
brane and a brane inflation with bulk scalar fields were discussed
respectively in \cite{aoyanagi06-brane-field,maeda00, himemoto01}.
In this paper, we suppose that in the $d$-dimensional bulk
spacetime, there is a general $m$-component $(m=d-1-p)$ scalar field
$\phi(x)=(\phi^1(x),\cdots,\phi^m(x))$. For general analysis, we
could expect this scalar field contains many topological and
dynamical information of the system. And by using a powerful tool -
the $\phi$-mapping topological tensor current theory
\cite{phi-mapping-original,apply-phi-mapping,phi-mapping-review}, we
show that this is indeed a fact. Actually, through a topological
tensor current constructed in terms of this field, we can not only
study the relationship of $p$-branes and the distribution of this
field, as well as some of their topological properties, but also
construct several new actions for $p$-branes, which govern in
general the dynamics of $p$-branes. Since our construction is in the
bulk spacetime and preserves many important symmetries naturally,
surely it will be of use to many applications.

The paper is organized as follows: In Sect.
\ref{phi-mapping-review}, we give a brief and quick review of the
$\phi$-mapping topological tensor current theory, during which we
also show that $p$-branes are generated at the zero points of the
field $\phi(x)$ and they are quantized in terms of their winding
numbers under the regular condition. In Sect. \ref{new-actions},
based on the tensor current just obtained, we construct three new
actions of $p$-branes, which are all defined in the bulk spacetime
and through a $\delta$-function can all reduce to the Nambu-Goto
action of $p$-branes. Also through introducing an auxiliary world
volume metric, we show that all actions just constructed change into
three other new forms. In Sect. \ref{properties-actions}, we discuss
some properties of the actions we constructed, such as they preserve
bulk Lorentz invariance and satisfy a self-dual condition. The
conclusion and some open problems are given in Sect.
\ref{conclusion}.

\section{Brief review of the $\phi-$mapping topological tensor
current theory}\label{phi-mapping-review}

The $\phi-$mapping topological tensor current theory, which was
proposed by Author Duan several years ago
\cite{phi-mapping-original}, has been found to be very powerful in
studying topological invariants and structures of physical systems,
for instances, it has been used to study the topological structures
of the Gauss-Bonnet-Chern theorem, the singular structures of the
London equation in superconductors, and the topological structures
of the cosmic strings as well as their bifurcations, \emph{etc}
\cite{apply-phi-mapping}. Recently, we found that besides the
topological properties of physical systems, this theory also
contains their dynamical properties. As it has been detailed
discussed in our previous papers \cite{phi-mapping-review}, here to
be self-contained, we only give it a brief and quick review. But
firstly, we give some notations used throughout this paper:
$g_{\mu\nu}$ with signature $(-1,1,\cdots,1)$ is the metric of the
$d$-dimensional bulk spacetime, where the greek indices $\mu,\nu$
run over $0,1,\cdots,d-1$. While,
\begin{equation}
h_{ab}=g_{\mu\nu}\frac{\partial x^\mu}{\partial\xi^a}\frac{\partial
x^\nu}{\partial\xi^b}
\end{equation}
is the induced metric in the $(p+1)$-dimensional world volume of
$p$-branes spanned by $(p+1)$ arbitrary parameters $\xi^a$, where
the latin indices $a,b$ run over $0,1,\cdots,p$. Also an auxiliary
world volume metric $\gamma_{ab}$ will be used in the next section.

By analogy with the discussion in \cite{phi-mapping-review}, one can
define a rank-$(p+1)$ topological tensor current as
\begin{equation}\label{current-definition}
j^{\mu_0\cdots\mu_p}
=\frac{1}{A(S^{m-1})(m-1)!}\epsilon^{\mu_0\cdots\mu_p\mu_{p+1}
 \cdots\mu_{d-1}}\epsilon_{\alpha_1\cdots\alpha_m}\partial_{\mu_
 {p+1}}\frac{\phi^{\alpha_1}}{\|\phi\|}\cdots\partial_{\mu_{d-1}}
 \frac{\phi^{\alpha_1}}{\|\phi\|}.
\end{equation}
where $\|\phi\|^2\!=\!\phi^\alpha\phi^\alpha$,
$\partial_\mu\!=\!\partial/\partial x^\mu$,
$A(S^{m-1})\!=\!2\pi^{m/2}/\Gamma(m/2)$ is the area of
$(m-1)-$dimensional unit sphere $S^{m-1}$, and the Levi-Civita
symbol $\epsilon^{\mu_0\cdots\mu_{d-1}}$ is defined to transform as
a tensor, which means that with all upper indices its components are
$\pm1/\sqrt{-g}$ and 0 while with all lower indices its components
are $\pm\sqrt{-g}$ and 0. Obviously this topological tensor current
is antisymmetric and identically conserved
\begin{equation}\label{conserve-j}
\nabla_{\mu_i}j^{\mu_0\cdots\mu_p}
=\frac{1}{\sqrt{-g}}\partial_{\mu_i}(\sqrt{-g}j^{\mu_0\cdots\mu_p})
=0,
\end{equation}
where the subscript $i\!=\!0,\cdots,d-1$. Then just as deduced in
\cite{phi-mapping-review}, one gets a compact $\delta-$function
structure of this tensor current rigorously
\begin{equation}\label{delta-j}
j^{\mu_0\cdots\mu_p}
  =\delta(\phi)J^{\mu_0\cdots\mu_p}\Big(\frac{\phi}{x}\Big),
\end{equation}
where the generalized Jacobian tensor is defined as
\begin{equation}\label{Jacobian-tensor}
\epsilon^{\alpha_1\cdots\alpha_m}J^{\mu_0\cdots\mu_p}
\Big(\frac{\phi}{x}\Big)
=\epsilon^{\mu_0\cdots\mu_p\mu_{p+1}\cdots\mu_{d-1}}\partial_
{\mu_{p+1}}\phi^{\alpha_1}\cdots\partial_{\mu_{d-1}}\phi^{\alpha_m},
\end{equation}
and which is nontrivial only when $\phi(x)\!=\!0$, or equivalently
\begin{equation}\label{phi=zero}
\phi^a(x)=0, \;\;\;\;a=1,\cdots,m.
\end{equation}
This is a natural result, as in the definition (\ref{delta-j}) of
the tensor current we have normalized the field. But in the
following we will see that this current structure actually involves
the total brane information of the system and indicates that all
branes are located at the zero points of $\phi(x)$. So we will focus
on these zero points.

Suppose that for the equations (\ref{phi=zero}), there are $K$
different solutions, namely the scalar fields $\phi(x)$ possess $K$
different zeros. According to the implicit function theorem, when
the regular condition of $\phi(x)$ for which the rank of the
Jacobian matrix $(\partial_\mu\phi^a)$ is $m$ satisfies, these
solutions can be expressed as
\begin{equation}                                                                
x^\mu=x^\mu_i(\xi^0,\cdots,\xi^p),  \;\;i=1,\cdots,K,
\end{equation}
where the subscript $i$ represents the $i$th solution which is a
$(p+1)$-dimensional submanifold spanned by the parameters
$\xi^a(a\!=\!0,\cdots,p)$ with the induced metric tensor $h_{ab}$
and called the $i$th singular submanifold $N_i$ in the bulk
spacetime. For each $N_i$ it can be proved that there exists a local
$m$-dimensional normal submanifold $M_i$ in the bulk spacetime
spanned by the parameters $\zeta^\alpha(\alpha\!=\!1,\cdots,m)$ with
the induced metric tensor $h'_{cd}\!=\!g_{\mu\nu}(\partial
x^\mu/\partial \zeta^c)(\partial x^\nu/\partial \zeta^d)$, which is
transversal to $N_i$ at the intersection point $p_i$. Then by virtue
of the implicit function theorem, at the regular point $p_i$, the
regular condition can be expressed explicitly as
\begin{equation}\label{regular-condition}
J\Big(\frac{\phi}{\zeta}\Big)
\equiv\frac{\partial(\phi^1,\cdots,\phi^m)}{\partial(\zeta^1,
\cdots,\zeta^m)}\neq0.
\end{equation}

Then according to the $\delta-$function theory, one can expand
$\delta(\phi)$ as
\begin{equation}\label{delta-phi}
\delta(\phi)=\sum_i\frac{\beta_i}{\Big|J\Big(\frac{\phi}
{\zeta}\Big)\Big|_{p_i}}\delta(N_i)
 =\sum_i\frac{\beta_i\eta_i}{J\Big(\frac{\phi}
{\zeta}\Big)|_{p_i}}\delta(N_i),
\end{equation}
where $\beta_i$ is a positive integer called the Hopf index of
$\phi-$mapping and $\eta_i\!=\!signJ(\phi/\zeta)|_{p_i}\!=\!\pm1$ is
the Brouwer degree, which are two topological invariant quantities.
$\delta(N_i)$ is the $\delta-$function on the singular submanifold
$N_i$ with the expression
\begin{equation}\label{delta-Ni}
\delta(N_i)=\!\int\!_{N_i}
\delta^d(x-x_i(\xi^0,\cdots,\xi^p))\sqrt{-h}d^{p+1}\xi,
\end{equation}
where $h$ is the determinant of the induced metric $h_{ab}$.
Substituting (\ref{delta-phi}) into (\ref{delta-j}), one gets the
final expansion form of the tensor current $j^{\mu_0\cdots\mu_p}$ on
the $K$ singular submanifolds
\begin{equation}\label{final-j1}                                                
j^{\mu_0\cdots\mu_p}
  =\sum_{i=1}^{K}\beta_i\eta_i\delta(N_i)
  \frac{J^{\mu_0\cdots\mu_p}\Big(\frac{\phi}{x}\Big)}
  {J\Big(\frac{\phi}{\zeta}\Big)|_{p_i}},
\end{equation}
or, in terms of parameters $y^{a}\!=\!(\xi^{a_0},\cdots,\xi^{a_p},
\zeta^{b_1},\cdots,\zeta^{b_m})$,
\begin{equation}\label{final-j2}                                                
j^{a_0\cdots a_p}
  =\sum_{i=1}^{K}\beta_i\eta_i\delta(N_i)
  \frac{J^{a_0\cdots a_p}\Big(\frac{\phi}{y}\Big)}
  {J\Big(\frac{\phi}{\zeta}\Big)|_{p_i}}.
\end{equation}

Now, we see that, if taking $\xi^0$ to be a timelike evolution
parameter $t$ and $\xi^1,\cdots,\xi^p$ spacelike parameters, as it
can be proved that $J^{\mu_0\cdots\mu_p}(\frac{\phi}{x})/J
(\frac{\phi}{\zeta})|_{p_i}$ or $J^{a_0\cdots a_p}(\frac{\phi}{y})
/J(\frac{\phi}{\zeta})|_{p_i}$ has the dimension of velocity, the
topological structures $j^{\mu_0\cdots\mu_p}$ (\ref{final-j1}) or
$j^{a_0\cdots a_p}$ (\ref{final-j2}) just represent $K$
$p$-dimensional isolated singular objects with charges
$\beta_i\eta_i$s moving in the bulk spacetime. These objects are
just $p$-branes, and the $(p+1)$-dimensional singular submanifolds
$N_i(i\!=\!1,\cdots,K)$ are their world volumes. Furthermore, we can
classify these $p$-branes in terms of their Brouwer degrees
$\eta_i$: a brane is called a $p$-brane if its $\eta\!>\!0$ and an
anti-$p$-brane if its $\eta\!<\!0$. Meanwhile, the product of the
Hopf index and Brouwer degree of a brane can also be denoted as
\begin{equation}
W=\beta\eta,
\end{equation}
which is an important topological quantum number - winding number -
of the brane, and this just shows that $p$-branes are quantized in
terms of their winding numbers.

\section{New actions of $p$-branes}\label{new-actions}

We have stated that the topological tensor current also contains the
dynamical properties of physical systems. Generally the dynamics of
$p$-branes are described by their actions. So we will study the
actions of $p$-branes and our interest is in constructing several
new actions. By using the tensor current $j^{\mu_0\cdots\mu_p}$
discussed in the above section, we can define an action of
$p$-branes as
\begin{equation}\label{action-1}
S_1=-\int d^dx\sqrt{-g}\sqrt{\frac{1}{(p+1)!}g_{\mu_0\nu_0}\cdots
g_{\mu_p\nu_p}j^{\mu_0\cdots\mu_p}j^{\nu_0\cdots\nu_p}},
\end{equation}
in which the corresponding Lagrangian density ${\cal L}$ is just a
generalization of Nielsen's Lagrangian \cite{nielsen73}. Then
considering the fact that on the singular submanifolds
$N_i(i=1,\cdots,K)$, $\phi(x)=0$, which leads to
$\partial_a\phi^\alpha(x)=\partial_\mu\phi^\alpha(x)\partial
x^\mu/\partial\xi^a\equiv0$, it can be proved that the generalized
Jacobian tensor (\ref{Jacobian-tensor}) and scalar
(\ref{regular-condition}) have the following relation
\begin{equation}\label{trans-Jacobi-tensor}
J^{\mu_0\cdots\mu_p}\Big(\frac{\phi}{x}\Big)
 =\epsilon^{a_0\cdots a_p}
  \frac{\partial x^{\mu_0}}{\partial\xi^{a_0}}\cdots
  \frac{\partial x^{\mu_p}}{\partial\xi^{a_p}}
  J\Big(\frac{\phi}{\zeta}\Big),
\end{equation}
so one gets
\begin{equation}
g_{\mu_0\nu_0}\cdots g_{\mu_p\nu_p}j^{\mu_0\cdots\mu_p}
j^{\nu_0\cdots\nu_p}
=(p+1)!\delta^2(\phi)J^2\Big(\frac{\phi}{\zeta}\Big),
\end{equation}
in which eq. (\ref{delta-j}) is used. Then the action $S_1$
simplifies to
\begin{equation}
S_1=-\int d^dx\sqrt{-g}
    \delta(\phi)\Big|J\Big(\frac{\phi}{\zeta}\Big)\Big|.
\end{equation}
Substituting (\ref{delta-phi}) and (\ref{delta-Ni}) into it yields
further that
\begin{equation}\label{action-1-reduce}
S_1=-\int d^dx\sqrt{-g}\sum_i\beta_i\int d^{p+1}\xi\sqrt{-h}
    \delta^d(x-x_i(\xi^0,\cdots,\xi^p))
   =\sum_iS_i,
\end{equation}
where
\begin{equation}\label{Nambu-Goto}
S_i=-\beta_i\int d^{p+1}\xi\sqrt{-h}
\end{equation}
is the action of the $i$th $p$-brane. Here it shows clearly that
besides a coefficient difference (or if we replace $\beta_i$ with
the tension of the $i$th $p$-brane $T_i$), this action is just the
Nambu-Goto action of the $i$th $p$-brane. In the following, we will
neglect this difference and just call (\ref{Nambu-Goto}) the
Nambu-Goto action of the $i$th $p$-brane. Also from
(\ref{action-1-reduce}), one sees that the action $S_1$ gives a
unified description of all $K$ $p$-branes.

For a further analysis of the action $S_1$ (\ref{action-1}), as
inside the square root, $j^{\mu_0\cdots\mu_p}$ still contains
derivatives, the relative calculation based on this action should be
very complicated. Actually, we can simplify it by introducing a
rank-$(p+1)$ symbol
\begin{equation}
k^{\mu_0\cdots\mu_p}=\epsilon^{a_0\cdots a_p}
  \frac{\partial x^{\mu_0}}{\partial\xi^{a_0}}\cdots
  \frac{\partial x^{\mu_p}}{\partial\xi^{a_p}},
\end{equation}
which satisfies
\[
k^{\mu_0\cdots\mu_p}k_{\mu_0\cdots\mu_p}=(p+1)!.
\]
Then a new action can be defined as
\begin{equation}\label{action-2}
S_2=-\int d^dx\sqrt{-g}\frac{1}{(p+1)!}k_{\mu_0\cdots\mu_p}
    j^{\mu_0\cdots\mu_p},
\end{equation}
where $k_{\mu_0\cdots\mu_p}$ now acts as a Lagrangian multiplier. By
using (\ref{trans-Jacobi-tensor}), a similar deduction suggests that
the action $S_2$ reduces to
\begin{equation}\label{action-2-reduce}
S_2=-\int d^dx\sqrt{-g}\delta(\phi)J\Big(\frac{\phi}{\zeta}\Big)
   =\sum_i\eta_iS_i,
\end{equation}
where $S_i$ is again the Nambu-Goto action (\ref{Nambu-Goto}) of the
$i$th $p$-brane.

If noting further that we have defined the Levi-Civita symbol
$\epsilon_{\mu_0\cdots\mu_{d-1}}$ as a tensor, which satisfies the
relation
\[
\epsilon_{\mu_0\cdots\mu_p}\frac{\partial x^{\mu_0}}
{\partial\xi^{a_0}}\cdots \frac{\partial x^{\mu_p}}
{\partial\xi^{a_p}} =\epsilon_{a_0\cdots a_p},
\]
(note as $N_i$ are submanifolds in the bulk spacetime, the inverse
of this relation is not correct), one gets that the action
(\ref{action-2}) has another equivalent but more simpler new form
\begin{equation}\label{action-3}
S_3=-\int d^dx\sqrt{-g}\frac{1}{(p+1)!}\epsilon_{\mu_0\cdots\mu_p}
    j^{\mu_0\cdots\mu_p},
\end{equation}
which can also reduce to eq. (\ref{action-2-reduce}).

Now we have constructed three different actions of $p$-branes in the
$d$-dimensional bulk spacetime by using the tensor current
$j^{\mu_0\cdots\mu_p}$, all of which can reduce to the Nambu-Goto
action of $p$-branes. But as is well known, actions can be
constructed in several different way, and one form or another will
be more useful for specific purposes. For example, the Nambu-Goto
action of $p$-branes has geometrical meaning and is intuitively easy
to understand, while the Polyakov action is very useful in the
covariant quantization. In the following of this section, we will
consider some other actions of $p$-branes in another way.

As a start, let us introduce a new independent auxiliary world
volume metric $\gamma_{ab}(\xi)$ of $p$-branes, which admits a
covariant gauge so that simplifies the analysis and allows a
covariant quantization. Henceforth `metric' will always mean
$\gamma_{ab}$ (unless we specify `induced'), and indices will always
be lowered and raised with this metric and its inverse
\cite{Polchinski-string-book}. The eq. (\ref{delta-Ni}) now changes
to the form
\begin{equation}
\delta(N_i)=\!\int\!_{N_i}
\delta^d(x-x_i(\xi^0,\cdots,\xi^p))\sqrt{-\gamma}d^{p+1}\xi,
\end{equation}
and the Levi-Civita symbol $\epsilon^{a_0\cdots a_p}$ has components
$\pm1/\sqrt{-\gamma}$ and 0, where $\gamma$ is the determinant of
$\gamma_{ab}$. Also the factor $\sqrt{-h}$ contained implicitly in
the actions $S_1$ (\ref{action-1}), $S_2$ (\ref{action-2}) and $S_3$
(\ref{action-3}) is replaced by $\sqrt{-\gamma}$, and so we obtained
another three new actions, which now are denoted as $S'_1$, $S'_2$,
$S'_3$ (Apparently, they have the same forms as $S_1$
(\ref{action-1}), $S_2$ (\ref{action-2}) and $S_3$
(\ref{action-3})). Just as the actions $S_1$, $S_2$, $S_3$ can
reduce to the Nambu-Goto action, we can also study the reduction of
the actions $S'_1$, $S'_2$, $S'_3$. But the results will show that
they correspond to different actions.

Let us first consider the action $S'_2$. As now one can prove
\begin{eqnarray}\label{gamma-h}
k_{\mu_0\cdots\mu_p}J^{\mu_0\cdots\mu_p}\Big(\frac{\phi}{x}\Big)
   \!&=&\!\epsilon^{a_0\cdots a_p}\epsilon^{b_0\cdots b_p}h_{a_0
          b_0}\cdots h_{a_pb_p}J\Big(\frac{\phi}{\zeta}\Big)
          \nonumber  \\
   \!&=&\!-(p+1)!h\gamma^{-1}J\Big(\frac{\phi}{\zeta}\Big),
\end{eqnarray}
in which the definition of the determinant of $h_{ab}$ is used, it
gives the action
\begin{equation}\label{action-2'-reduce}
S'_2\rightarrow-\int d^dx\sqrt{-g}
(-h\gamma^{-1})\delta(\phi)J\Big(\frac{\phi}{\zeta}\Big)
=-\sum_i\eta_iS_i,
\end{equation}
where
\begin{equation}
S_i=-\beta_i\int d^{p+1}\xi\sqrt{-\gamma}h\gamma^{-1}.
\end{equation}
This is just the Schild type (null) $p$-brane action
\cite{nieto-01}. So that when replace $\sqrt{-h}$ by
$\sqrt{-\gamma}$, through the function $\delta(\phi)$, the action
$S'_2$ reduce to the Schild type action of $p$-branes, not the
Nambu-Goto action any more.

Secondly, we consider $S'_1$. From the properties of the Levi-Civita
symbol
\[
\epsilon^{a_0\cdots a_p}\epsilon_{c_0\cdots c_p}
 =\delta^{a_0\cdots a_p}_{c_0\cdots c_p},\;\;\;
 \epsilon^{b_0\cdots b_p}=\gamma^{b_0c_0}\cdots
 \gamma^{b_pc_p}\epsilon_{c_0\cdots c_p},
\]
one can get
\begin{eqnarray}
g_{\mu_0\nu_0}\cdots g_{\mu_p\nu_p}J^{\mu_0\cdots\mu_p}\Big(\frac
          {\phi}{x}\Big)J^{\nu_0\cdots\nu_p}\Big(\frac{\phi}{x}\Big)
   \!&=&\!\epsilon^{a_0\cdots a_p}\epsilon^{b_0\cdots b_p}
          h_{a_0b_0}\cdots h_{a_pb_p}J^2\Big(\frac{\phi}{\zeta}
          \Big)  \nonumber  \\
   \!&=&\!(p+1)!(\gamma^{ab}h_{ab})^{p+1}J^2\Big(\frac{\phi}{\zeta}
          \Big).
\end{eqnarray}
So that the action reduces to
\begin{equation}\label{action-1'-reduce}
S'_1\rightarrow-\int d^dx\sqrt{-g}(\gamma^{ab}h_{ab})^{\frac
 {p+1}{2}}\delta(\phi)\Big|J\Big(\frac{\phi}{\zeta}\Big)\Big|
 =\sum_iS_i
\end{equation}
where
\begin{equation}
S_i=-\beta_i\int
d^{p+1}\xi\sqrt{-\gamma}(\gamma^{ab}h_{ab})^{\frac{p+1}{2}}.
\end{equation}
This action is just the so-called second Weyl-invariant $p$-brane
action discussed in \cite{miao03-parent-actions,nieto-01}, which has
an important property - Weyl invariance, namely, it preserves
invariance under the following Weyl transformation
\begin{eqnarray}
x^\mu(\xi)\!&\rightarrow&\!x^\mu(\xi), \nonumber  \\
\gamma_{ab}(\xi)\!&\rightarrow&\!
exp\big(2\omega(\xi)\big)\gamma_{ab}(\xi), \nonumber
\end{eqnarray}
where $\omega(\xi)$ is an arbitrary real function of the world
volume parameters of $p$-branes. Further, based on this action,
through a parent action, the first Weyl-invariant action can be
obtained \cite{miao03-parent-actions,nieto-01}. And as the
Weyl-invariance of actions plays a central role in the
string/M-theory, the action $S_2$ is surely worth further studies.

Still for $S'_1$, if using the relation
\begin{eqnarray}
g_{\mu_0\nu_0}\cdots g_{\mu_p\nu_p}J^{\mu_0\cdots\mu_p}\Big(\frac
          {\phi}{x}\Big)J^{\nu_0\cdots\nu_p}\Big(\frac{\phi}{x}\Big)
   \!&=&\!\epsilon^{a_0\cdots a_p}\epsilon^{b_0\cdots b_p}
          h_{a_0b_0}\cdots h_{a_pb_p}J^2\Big(\frac{\phi}{\zeta}
          \Big)  \nonumber  \\
   \!&=&\!-(p+1)!h\gamma^{-1}J^2\Big(\frac{\phi}{\zeta}\Big).
\end{eqnarray}
which is similar to (\ref{gamma-h}), as
\begin{equation}
\int d^{p+1}\xi\sqrt{-\gamma}\sqrt{h\gamma^{-1}}
 =\int d^{p+1}\xi\sqrt{-h},
\end{equation}
it reduces to the Nambu-Goto action of $p$-branes again. And this
shows the classical equivalence of $S'_1$ and $S_1$.

Finally, for the action $S'_3$, it is easy to show that it will not
reduce to new actions but still gives the Nambu-Goto action through
$\delta(\phi)$, and we will not give a repeat deduction.

\section{Properties of the new actions}\label{properties-actions}

Generally the topological properties of $p$-branes have been shown
in Sect. \ref{phi-mapping-review}. So in this section, we mainly
study their dynamical properties, namely the properties of their
actions. We see that though through a $\delta$-function
$\delta(\phi)$, the actions $S_1$ (\ref{action-1}), $S_2$
(\ref{action-2}) and $S_3$ (\ref{action-3}) can all reduce to the
Nambu-Goto action (and $S'_1$, $S'_2$, $S'_3$ reduce to some other
simple actions), they have much richer structures. Firstly, from
eqs. (\ref{action-1-reduce}), (\ref{action-2-reduce}),
(\ref{action-2'-reduce}) and (\ref{action-1'-reduce}), each of the
actions gives a unified description of all $K$ $p$-branes, which is
equivalent to say that each action describes the total dynamical
properties of the brane system. Also compared to the Nambu-Goto
action which is defined on the world volume of $p$-branes, they all
defined in the bulk spacetime naturally, and so that all preserve
manifestly the full spacetime Lorentz invariance.

Then by using the parent action method we consider another important
property of the actions (\ref{action-1}), (\ref{action-2}), and
(\ref{action-3}), and take the action $S_2$ (\ref{action-2}) as an
example. The parent action approach was originally proposed to
establish the equivalence or so-called \emph{duality} between the
Abelian self-dual and Maxwell-Chern-Simons models in
(2+1)-dimensional spacetime, at the level of the Lagrangian instead
of equations of motion, and recently it has been developed in many
directions and become a powerful tool to display the duality
relations of different actions
\cite{parent-action,miao03-parent-actions}. According to this
method, we introduce two rank-$(p+1)$ tensor fields
$\Lambda_{\mu_0\cdots\mu_p}$ and $p^{\mu_0\cdots\mu_p}$, and write
down a parent action of $S_2$
\begin{equation}\label{parent-action-2}
S^{parent}_2=-\int d^dx\sqrt{-g}\frac{1}{(p+1)!}
   [k_{\mu_0\cdots\mu_p}p^{\mu_0\cdots\mu_p}
   +\Lambda_{\mu_0\cdots\mu_p}(p^{\mu_0\cdots\mu_p}-
    j^{\mu_0\cdots\mu_p})],
\end{equation}
where $\Lambda_{\mu_0\cdots\mu_p}$ and $p^{\mu_0\cdots\mu_p}$ are
treated as two independent auxiliary fields. Now varying the parent
action (\ref{parent-action-2}) with respect to
$\Lambda_{\mu_0\cdots\mu_p}$ gives the relation
\begin{equation}
p^{\mu_0\cdots\mu_p}=j^{\mu_0\cdots\mu_p}.
\end{equation}
Then together with this relation, the action (\ref{parent-action-2})
turns back to the original action (\ref{action-2}), and this shows
the classical equivalence between the two actions. Meanwhile, if
varying (\ref{parent-action-2}) with respect to
$p^{\mu_0\cdots\mu_p}$, it leads to the relation
\begin{equation}
\Lambda_{\mu_0\cdots\mu_p}=-k_{\mu_0\cdots\mu_p}.
\end{equation}
Substituting it into (\ref{parent-action-2}) yields again the
original action (\ref{action-2}). Thus we get that the action $S_2$
(\ref{action-2}) is actually a \emph{self-dual} action. The same
result also applies to the actions $S_1$ (\ref{action-1}) and $S_2$
(\ref{action-3}), and the parent actions are, respectively
\begin{eqnarray}
S^{parent}_1\!&=&\!-\int d^dx\sqrt{-g}
   \left[\sqrt{\frac{1}{(p+1)!}p^{\mu_0\cdots\mu_p}
   p_{\mu_0\cdots\mu_p}}+\Lambda_{\mu_0\cdots\mu_p}
   (p^{\mu_0\cdots\mu_p}-j^{\mu_0\cdots\mu_p})\right],
   \label{parent-action-1}\\
S^{parent}_3\!&=&\!-\int d^dx\sqrt{-g}\frac{1}{(p+1)!}
   [\epsilon_{\mu_0\cdots\mu_p}p^{\mu_0\cdots\mu_p}
   +\Lambda_{\mu_0\cdots\mu_p}(p^{\mu_0\cdots\mu_p}-
    j^{\mu_0\cdots\mu_p})].
\end{eqnarray}
(Note that to show the self-duality of $S_1$, it is a little more
complicated than $S_2$, which means that after substituting the
relation obtained by varying $p^{\mu_0\cdots\mu_p}$ into the parent
action (\ref{parent-action-1}) one needs a further variation in
terms of $p^{\mu_0\cdots\mu_p}$). Moreover, as apparently the
actions $S'_1$, $S'_2$, $S'_3$ have the same forms as $S_1$, $S_2$,
$S_3$, the above discussion also apply to them, which is to say that
$S'_1$, $S'_2$, $S'_3$ are also self-dual actions.

\section{Conclusion}\label{conclusion}

In this paper, through a brief review of the $\phi-$mapping
topological tensor current theory, we first discuss some topological
properties of $p$-branes, and the main results are that all
$p$-branes are generated at the zero points of the field $\phi(x)$
and they are quantized in terms of their winding numbers. Then we
discuss the construction of three new actions $S_1$, $S_2$, $S_3$ of
$p$-branes. We show that based on the results we obtained in the
$\phi$-mapping theory and through the $\delta$-function
$\delta(\phi)$, all these three actions can be reduced to the basic
Nambu-Goto action of $p$-branes. Further, through introducing an
auxiliary world volume metric $\gamma_{ab}$, which is used to raise
and lower indices, all three actions take new forms $S'_1$, $S'_2$,
$S'_3$, which then can be reduced to the second Weyl-invariant, the
Schild type and the Nambu-Goto actions, respectively. Finally, we
discuss some properties of the actions that we construct. We show
that compared to the Nambu-Goto action, all these actions are
defined naturally in the bulk spacetime, and so preserve the full
spacetime Lorentz invariance. Also, by using the parent action
approach, we show that all actions we construct have an important
property: they are all self-dual type actions.

Further, we have seen that during the construction of new actions,
the Nambu-Goto action plays an important role. But the
non-polynomial form of this action makes it difficult for canonical
analysis and quantization. Other reduced actions from $S'_1$ and
$S'_2$ also have the same or worse drawbacks. Can we construct some
other actions in terms of our method, which can reduce to actions
that are convenient to analyze, for example the Polyakov action or
the so-called first Weyl-invariant action of $p$-branes? Also,we
need to carry further analysis of the actions we have constructed,
such as the canonical analysis, the supersymmetric extension,
\emph{etc}. They are our further works.

\section*{Acknowledgments}                                               

One of the authors ZBC is indebted to Dr. Y. X. Liu for his much
help. This work was supported by the National Natural Science
Foundation and the Doctor Education Fund of Educational Department
of the People’s Republic of China.


\end{document}